# Quantum interference in second-harmonic generation from monolayer WSe₂

Kai-Qiang Lin[1,2]*, Sebastian Bange[1], John M. Lupton[1]*

[1]Institut für Experimentelle und Angewandte Physik,

Universität Regensburg, 93053 Regensburg, Germany.

[2]Collaborative Innovation Center of Chemistry for Energy Materials, College of

Chemistry and Chemical Engineering, Xiamen University, 361005 Xiamen, China.

A hallmark of wave-matter duality is the emergence of quantum-interference phenomena when an electronic transition follows different trajectories. Such interference results in asymmetric absorption lines such as Fano resonances[1], and gives rise to secondary effects like electromagnetically induced transparency (EIT) when multiple optical transitions are pumped[2-5]. Few solid-state systems show quantum interference and EIT[5-11], with quantum-well intersubband transitions in the IR[12,13] offering the most promising avenue to date to devices exploiting optical gain without inversion[14,15]. Quantum interference is usually hampered by inhomogeneous broadening of electronic transitions, making it challenging to achieve in solids at visible wavelengths and elevated temperatures. However, disorder effects can be mitigated by raising the oscillator strength of atom-like electronic transitions – excitons – which arise in monolayers of transition-metal dichalcogenides (TMDCs)[16,17]. Quantum interference, probed by second-harmonic generation (SHG)[18,19], emerges in $WSe_2$, without a cavity, splitting the SHG spectrum. The splitting exhibits spectral anticrossing behaviour, and is related to the number of Rabi flops the strongly driven system undergoes. The SHG power-law exponent deviates strongly from the canonical value of 2, showing a Fano-like wavelength dependence which is retained at room temperature. The work opens opportunities in solid-state quantum-nonlinear optics for optical mixing, gain without inversion and quantum-information processing.

We consider the wavelength dependence of SHG from monolayer $WSe_2$ crystals as sketched in Fig. 1a. Under excitation at 724 nm with pulses of 80 fs length at 80 MHz repetition rate, Fig. 1b reveals two peaks in the SHG spectrum, at 360 nm and 364 nm, along with a broad feature at 581 nm. The latter is attributed to upconversion photoluminescence (UPL)[20]. As plotted in the inset, SHG and UPL are distinguished by the variation in emission intensity copolarized with the laser as the polarization plane of the driving field is rotated with respect to the crystal[18]. The luminescence (red) is independent of crystal orientation whereas the two SHG peaks (blue, yellow) exhibit the characteristic six-fold symmetry expected from the three-fold rotational crystal symmetry[18]. The spectral SHG dip depends on excitation wavelength $\lambda_{ex}$, as shown for normalized SHG spectra in the left panel of Fig. 1c. The two peaks display anticrossing behaviour instead of tracking the excitation wavelength, with the usual single-peak Gaussian shape of the transform-limited SHG spectrum returning at long and short pump wavelengths. The spectral dip coincides with a minimum in integrated SHG efficiency as a function of $\lambda_{ex}$, as shown in Fig. S1, implying that part of the SHG spectrum is suppressed. No such suppression is seen in the UPL excitation spectrum in Fig. S1.

The narrow-band attenuation in the broad SHG spectrum is reminiscent of EIT, which is commonly probed in linear transmission[3,5], and is interpreted here in terms of quantum-interference effects disturbing resonant excitonic enhancement of SHG. As detailed in the *Supplementary Information* (Figs. S2, S3), after discarding alternative interpretations, we model the nonlinear scattering response in terms of a ladder-type three-level system typical of EIT with degenerate control and probe beams. Fig. 1d shows a three-level scheme involving two dipole-allowed transitions. The PL spectrum of the sample in Fig. S1 reveals the narrow "A" exciton feature at 720 nm, which we identify with the $|1\rangle \rightarrow |2\rangle$ transition. We propose the existence of an additional state $|3\rangle$ which supports a $|2\rangle \rightarrow$ $|3\rangle$ transition around 724 nm. While this transition is close to resonant with the driving field, it does not contribute to SHG enhancement due to vanishing $|1\rangle \rightarrow |3\rangle$ oscillator strength. We propose that the dipole-allowed $|2\rangle \rightarrow |3\rangle$ transition occurs between the lowest-energy and a higher-energy conduction band as has been identified in realistic band structure calculations[21] (Fig. S4). Dressing of level $|2\rangle$ under strong driving of the $|2\rangle \rightarrow |3\rangle$ transition is thus likely linked to the appearance of Floquet-Bloch states and photoinduced band inversions[21]. State dressing is then directly observable in the overall strength of resonant SHG and can be viewed as quantum interference between the allowed transitions $|1\rangle \rightarrow |2\rangle$ and $|3\rangle \rightarrow |2\rangle$[13]. The excitonic transition therefore mixes with the band continuum, a phenomenon associated with the elusive non-linear Fano effect[22]. With these simple assumptions and realistic parameters summarised in Table S1 we can simulate the SHG spectrum within a density-matrix formalism, replicating the SHG dip in the $\lambda_{ex}$-dependent spectrum in the right of Fig. 1c.

We test the model by examining predictions with respect to width and position of the dip. The model stipulates that the SHG dip width is not determined by the linewidth of the excitonic transitions but is linked to laser pulse length and intensity. We modify the excitonic linewidth by comparing bare monolayer crystals to those encapsulated between layers of hBN[23] in Fig. S6. Without encapsulation, the A exciton (state $|2\rangle$) transition shows a width of 17.6 meV in PL at 5 K, compared to 6.3 meV for the encapsulated sample. In both cases, the width of the dip is 23.2 meV. Next, we vary the incident field. The density matrix dynamics in Fig. 2b reveal Rabi flopping between the populations of states $|2\rangle$ and $|3\rangle$, i.e. $\rho_{22}(t)$ (red line) and $\rho_{33}(t)$ (blue line). The calculations in panel a-c, parameterized following Table S1, predict that either an increase in pulse energy or in pulse length raises the number of Rabi flops the system undergoes on the timescale of the excitation pulse. This effect determines the number of spectral dips that appear in the calculated SHG spectra in Fig. 2d-f (dark red lines). Such a trifurcation of the SHG spectrum is indeed observed in experiment when increasing the pulse energy from 3.3 pJ to 50 pJ for laser pulses of 80 fs length. The pulse energy required in experiment is higher than what the simulation predicts, presumably due to the onset of many-body effects in this regime of strong pumping. Simulated SHG spectra can be brought into agreement with experiment as shown by the pale-red line in Fig. 2d, by taking into account qualitatively the impact of many-body interactions: a reduction of transition dipole moments and an increase in decoherence rate[27]. Details of these adjustments and estimates of excitation density are discussed in Figs. S7, S14. Without changing the pulse energy, we obtain the double-dip SHG spectrum in Fig. 2i by raising the pulse length from 80 fs to 140 fs. The evolution of SHG spectra with excitation wavelength illustrates further agreement between simulation and experiment for the three-peak spectra (Fig. S8).

The effective excitation power driving optical transitions can be increased by exploiting the pseudospin valleys formed around the $\pm K$ points in momentum space, which enable valley-selective optical pumping[28]. Circularly polarized excitation drives the excitonic transition in a single valley, so that the SHG radiation has opposite handedness with respect to the pump (Fig. S9). When compared to excitation with linearly polarized light of the same power, the threshold pump power required to achieve doubly split SHG spectra under circularly polarized excitation is reduced by an order of magnitude (Fig. S9). This reduction is consistent with the notion of an increase in the effective valley-selective pump strength, with the laser field driving quantum interference close to one of the two K valleys.

Besides altering excitation power to probe the quantum interference phenomenon, sample temperature offers a parameter to tune the influence of decoherence and inhomogeneous broadening. While these material-dependent effects are not easily accounted for in the model, an increase in temperature is expected to raise decoherence rates, lowering the strength of light-matter coupling and therefore bleaching the SHG dip as simulated in Fig. S10. Figure 3 shows the evolution of SHG anticrossing with temperature. As temperature increases, the SHG dip disappears. Above 200 K the emission wavelength of SHG shows the usual linear dependence on excitation wavelength ($\lambda_{\mathrm{em}} = \lambda_{\mathrm{ex}}/2$).

Even though the SHG dip appears suppressed at elevated temperatures, it is conceivable that the excitonic pathways for quantum interference may remain in place. We explore this possibility through an analysis of the excitation-power dependence of SHG in Fig. S11, which typically follows a power law $I_{\mathrm{SHG}}(\lambda_{\mathrm{em}}) \propto I_{\mathrm{ex}}^{p(\lambda_{\mathrm{em}})}$. The exponent $p(\lambda_{\mathrm{em}})$ extracted in Fig. 4a varies across the spectrum, dropping to 0.6 and rising up to 3.0 in the region of bifurcation. The asymmetric shape of $p(\lambda_{\mathrm{em}})$ is reminiscent of the Fano function, superimposed in red. We interpret the fact that the spectral dependence of the SHG power-law exponent is Fano-like as another signature of the quantum interference effect underlying the SHG. Interestingly, the sub-parabolic behaviour of $p(\lambda_{em})$ is retained at elevated temperatures (Fig. S11b), even as the SHG dip disappears (Fig. 3). Since the spectral region of the interference is too broad to be fully accounted for by $p(\lambda_{em})$, we consider the pump-intensity dependence of spectrally integrated SHG $\int I_{SHG}(\lambda_{em}, \lambda_{ex}) \, d\lambda_{em} \propto I_{ex}^{p(\lambda_{ex})}$ as a function of excitation wavelength. This analysis yields the power-law exponent of SHG in excitation, $p(\lambda_{ex})$. Examples of the pump-intensity dependence of SHG for two different $\lambda_{ex}$ are given in Fig. S12. In Fig. 4b we compare $p(\lambda_{ex})$ to $p(\lambda_{em})$, taken from Fig. 4a. The exponents differ in terms of magnitude, since $p(\lambda_{em})$ averages over the spectral width of the SHG, but the exponents otherwise show the same dispersive functionality. Figure S12 plots a calculated spectrum of $p(\lambda_{ex})$ which is in good qualitative agreement with experiment. We conclude that both the SHG spectrum and the dependence of integrated SHG intensity on excitation wavelength probe the same quantum-interference process. The spectral dependence of the SHG power-law exponent can therefore be used as a metric for further exploration of the model. Finally, we consider the temperature dependence of the $p(\lambda_{ex})$ spectrum in Fig. 4c. The Fano-like spectral evolution of sub- and super-parabolic responses of the SHG is retained even at room temperature, suggesting that quantum interference may persist under these conditions.

The extraordinary electronic structure of monolayer TMDCs dramatically enhances the strength of light-matter coupling, leading to valley-selective quantum interference in surface-SHG in the absence of an external cavity. We stress this point since almost identical features are resolved in monolayers deposited on $SiO_2$/Si substrates and on smooth gold films, which suppress wave guiding (see Fig. S13 for a complete discussion). The observation of spectroscopic signatures of quantum interference associated with a ladder-type three-level system implies that monolayer TMDCs should permit for inversionless gain given suitable electronic resonances, which can be tuned by temperature, dielectric screening in sandwich structures, and electrical gating. The persistence at room temperature of signatures of Fano-like dispersion in the spectral dependence of the SHG power-law exponent raises the question of the possibility of designing optoelectronic devices exploiting otherwise elusive quantum-interference phenomena under ambient conditions. Fano-like resonances have recently attracted attention in the field of plasmonics[29], where they spectrally narrow the inherently broad resonant response of nanoparticle plasmons, potentially increasing spectral selectivity in nanoscale sensors. The emergence of such resonances in monolayer TMDC optics may serve to enhance resolution in surface-sensitive processes, enable the design of new ultrafast pulse and phase-shaping tools, and could offer a route to atomically thin optical-parametric amplifiers exploiting resonantly enhanced four-wave mixing.

**Methods**

**Sample preparation.** WSe$_2$ monolayers and thin layers of hBN were obtained through mechanical exfoliation from bulk crystals (HQ Graphene) onto commercial PDMS films (Gel-Pak, Gel-film® X4) with blue Nitto tape (Nitto Denko, SPV 224P)[30]. We deposited either bare WSe$_2$ layers or WSe$_2$ encapsulated on either side with hBN on a silicon wafer with 300 nm of thermal oxide on top. The stamp transfer was performed using an optical microscope combined with translation stages to allow precise placement of the WSe$_2$ monolayer between two hBN layers. The substrate temperature was controlled by a small heating stage based on a Peltier module, and was generally 65°C before the layer on the PDMS stamp was attached to a bare silicon substrate or the layer already present on the wafer. Optical microscope images of the hBN-encapsulated monolayer WSe$_2$ sample discussed in the main text are shown in Fig. S16 for different stages of fabrication.

**Optical spectroscopy.**

i) SHG and UPL

As sketched in Fig. S17, a Ti:sapphire laser with 80 fs pulse length (Newport Spectra-Physics, Mai Tai XF, 80 MHz repetition rate) or 140 fs pulse length (Coherent, Chameleon Ultra II, 80 MHz repetition rate) was focused through a 0.6 numerical aperture microscope objective (Olympus, LUCPLFLN 40×, with a coverslip correction capability of up to 2 mm) onto the WSe$_2$ monolayer sample mounted under vacuum on the cold finger of a helium microscope cryostat (Janis, ST-500, using a 1.57 mm thick fused-silica window). The diameter of the laser spot was estimated to be 1.9 μm. The incident laser power was set by an electrically controlled neutral-density filter wheel together with a power meter. A 10:90 non-polarizing cube beam splitter (Thorlabs, BS025) was generally used to separate incident path and signal detection path. To achieve higher pump power and detection efficiency, a 647 nm single-edge short-pass dichroic beam splitter (Semrock, FF647-SDi01) was used for measurements in Fig. 2, Fig. S7, Fig. S8, and Fig. S14. The signal was collected by the same objective, dispersed in a spectrometer (Princeton Instruments, Acton SP2300) with two gratings installed (150 grooves/mm and 1200 grooves/mm) and read out by a CCD camera (Princeton Instruments, PIXIS 100). The lower-resolution grating was used to capture SHG and UPL simultaneously. The finer grating was used to obtain high-resolution SHG spectra. A sCMOS camera (Hamamatsu, ORCA-Flash 4.0) was used to acquire sample images.

ii) Polarization-dependent SHG and UPL measurements

In order to measure the excitation polarization-angle dependence of copolarized SHG and UPL (Fig. 1b of the main text), the polarization of the horizontally polarized laser beam was rotated to the excitation-polarization plane by a superachromatic half-wave plate (Thorlabs, SAHWP05M-700) inserted between a non-polarizing beam splitter (~10% reflectivity) and the objective. The SHG and UPL signals were collected by the objective, and passed again through the half-wave plate. The light was spectrally filtered by a 680 nm short-pass edge filter (Semrock, FF01-680SP) and detected in the horizontal polarization plane by inserting a corresponding linear polarizer directly after the emission filter. The half-wave plate was rotated by a stepping motor in 1° steps from 0 to 360° to continuously vary the laser polarization with respect to the fixed crystal orientation.

Multiple rotations were performed to confirm that the signal did not degrade during the measurement process.

### iii) Spectrally resolved power-law exponent of SHG in emission $p(\lambda_{em})$

To obtain the spectrally resolved power-law exponent of SHG as shown in the bottom panel of Fig. 4a, we first performed high-resolution SHG intensity measurements as a function of pump power (Fig. S11a). The pump power was measured between the beam splitter and the objective. Power levels (corresponding energy per pulse) of 370 µW (4.6 pJ), 260 µW (3.3 pJ), 180 µW (2.2 pJ), 130 µW (1.7 pJ), 90 µW (1.1 pJ) and 370 µW (4.6 pJ) were used for a single cycle of measurements. The laser power was returned to 370 µW at the end of the sequence to confirm that no defocusing or degradation arose. By plotting the power-dependent SHG intensity for each emission wavelength on a double-logarithmic scale, we obtain the spectrally resolved SHG power-law exponent through a linear fit of the six points acquired for each emission wavelength.

### iv) Excitation-wavelength-dependent power-law exponent of integrated SHG $p(\lambda_{ex})$

To measure the power-law exponent of the integrated SHG spectrum as shown in Fig. 4b,c, we performed power-dependent integrated SHG intensity measurements for five different excitation powers (pulse energies) – 370 µW (4.6 pJ), 260 µW (3.3 pJ), 180 µW (2.2 pJ), 130 µW (1.7 pJ) and 90 µW (1.1 pJ) – in sweeps up and down in power for each excitation wavelength. By plotting the integrated area of the SHG spectrum as a function of excitation power on a double-logarithmic scale as shown in Fig. S12a, we obtain the SHG power-law exponent through a linear fit for each excitation wavelength.

v) Valley-resolved SHG measurements

To generate circularly polarized excitation, a Berek-type variable-wave plate (Newport Spectra-Physics) was inserted in the laser beam path. The signal helicity was analyzed by means of a rotatable superachromatic quarter-wave plate (B. Halle Nachfl. GmbH) together with a laser-quality calcite Glan-Taylor polarizer inserted after a 680 nm short-pass edge filter (Semrock FF01-680SP) in the detection path.

vi) Photoluminescence (PL)

The 488 nm line of an argon-ion laser (Spectra-Physics, 2045E) was used to excite the one-photon PL of monolayer $WSe_2$. After passing through a 488 nm long-pass edge filter (Semrock, LP02-488RU), the PL signal was dispersed by a 150 grooves/mm grating and recorded by the CCD camera. PL was also measured with the Chameleon femtosecond laser set to 680 nm. In this case, a 700 nm short-pass edge filter (Thorlabs, FES0700) was placed in the excitation path and a 700 nm long-pass edge filter (Thorlabs, FEL0700) was used to suppress the laser line in the detection path.

**Fano resonance line shape.** We superimposed the spectrally resolved power-law exponent in Fig. 4a with an asymmetric Fano line shape[1] of the form

$$p = p_0 + \frac{H(q\Gamma_{res}/2 + E - E_{res})^2}{(\Gamma_{res}/2)^2 + (E - E_{res})^2}$$

where $q$ is the Fano parameter, $E_{res}$ is the resonance energy, $\Gamma_{res}$ is the resonance linewidth and $H$ is the amplitude of the resonance. The best agreement between the measured data is found for the parameter set $q = -1.7$, $E_{res} = 3.42$ eV (362 nm), $p_0 = 0.71$, $H = 0.56$, and $\Gamma_{res} = 14.56$ meV.

**Acknowledgements**

The authors thank A. Chernikov, R. Huber, T. Korn, F. Langer, P. Nagler, A. Kormányos, and B. Ren for helpful discussions, and R. Martin for assistance with sample preparation. Financial support is gratefully acknowledged from the German Science Foundation through SFB 1277 project B3.

**Author contributions**

K. L. conceived and performed the experiments with the support of S. B. S. B. wrote the simulation codes and carried out simulations with K. L. K. L., S. B. and J. M. L. analyzed the data and wrote the paper.

**Author information**

The authors declare that they have no competing financial interests.

Correspondence and requests for materials should be addressed to K.L. (kaiqiang.lin@ur.de) or J.M.L. (john.lupton@ur.de).

**References**

1.      Fano, U. Effects of configuration interaction on intensities and phase shifts. *Phys. Rev.* **124**, 1866-1878 (1961).

2.      Harris, S. E., Field, J. E. & Imamoğlu, A. Nonlinear optical processes using electromagnetically induced transparency. *Phys. Rev. Lett.* **64**, 1107-1110 (1990).

3.      Harris, S. E. Electromagnetically induced transparency. *Phys. Today* **50**, 36-42 (1997).

4.      Lukin, M. D. & Imamoğlu, A. Controlling photons using electromagnetically induced transparency. *Nature* **413**, 273-276 (2001).

5.      Fleischhauer, M., Imamoglu, A. & Marangos, J. P. Electromagnetically induced transparency: optics in coherent media. *Rev. Mod. Phys.* **77**, 633-673 (2005).

6.      Ham, B. S., Hemmer, P. R. & Shahriar, M. S. Efficient electromagnetically induced transparency in a rare-earth doped crystal. *Opt. Comm.* **144**, 227-230 (1997).

7.      Wei, C. & Manson, N. B. Observation of the dynamic Stark effect on electromagnetically induced transparency. *Phys. Rev. A* **60**, 2540-2546 (1999).

8.      Turukhin, A. V. *et al.* Observation of ultraslow and stored light pulses in a solid. *Phys. Rev. Lett.* **88**, 023602 (2001).

9.      Phillips, M. C. *et al.* Electromagnetically induced transparency in semiconductors via biexciton coherence. *Phys. Rev. Lett.* **91**, 183602 (2003).

10.     Longdell, J. J., Fraval, E., Sellars, M. J. & Manson, N. B. Stopped light with storage times greater than one second using electromagnetically induced transparency in a solid. *Phys. Rev. Lett.* **95**, 063601 (2005).

11.    Abdumalikov, A. A. *et al.* Electromagnetically induced transparency on a single artificial atom. *Phys. Rev. Lett.* **104**, 193601 (2010).

12.    Faist, J., Capasso, F., Sirtori, C., West, K. W. & Pfeiffer, L. N. Controlling the sign of quantum interference by tunnelling from quantum wells. *Nature* **390**, 589-591 (1997).

13.    Serapiglia, G. B., Paspalakis, E., Sirtori, C., Vodopyanov, K. L. & Phillips, C. C. Laser-induced quantum coherence in a semiconductor quantum well. *Phys. Rev. Lett.* **84**, 1019-1022 (2000).

14.    Scully, M. O. & Fleischhauer, M. Lasers without inversion. *Science* **263**, 337-338 (1994).

15.    Frogley, M. D., Dynes, J. F., Beck, M., Faist, J. & Phillips, C. C. Gain without inversion in semiconductor nanostructures. *Nat. Mater.* **5**, 175-178 (2006).

16.    Chernikov, A. *et al.* Exciton binding energy and nonhydrogenic Rydberg series in monolayer $WS_2$. *Phys. Rev. Lett.* **113**, 076802 (2014).

17.    He, K. L. *et al.* Tightly bound excitons in monolayer $WSe_2$. *Phys. Rev. Lett.* **113**, 026803 (2014).

18.    Seyler, K. L. *et al.* Electrical control of second-harmonic generation in a $WSe_2$ monolayer transistor. *Nat. Nanotechnol.* **10**, 407-411 (2015).

19.    Wang, G. *et al.* Giant enhancement of the optical second-harmonic emission of $WSe_2$ monolayers by laser excitation at exciton resonances. *Phys. Rev. Lett.* **114**, 097403 (2015).

20.    Manca, M. *et al.* Enabling valley selective exciton scattering in monolayer $WSe_2$ through upconversion. *Nat. Commun.* **8**, 14927 (2017).

21. Claassen, M., Jia, C., Moritz, B. & Devereaux, T. P. All-optical materials design of chiral edge modes in transition-metal dichalcogenides. *Nat. Commun.* **7**, 13074 (2016).

22. Kroner, M. *et al.* The nonlinear Fano effect. *Nature* **451**, 311-314 (2008).

23. Cadiz, F. *et al.* Excitonic linewidth approaching the homogeneous limit in $MoS_2$-based van der Waals heterostructures. *Phys. Rev. X* **7**, 021026 (2017).

24. Sun, D. *et al.* Observation of rapid exciton–exciton annihilation in monolayer molybdenum disulfide. *Nano Lett.* **14**, 5625-5629 (2014).

25. Chernikov, A., Ruppert, C., Hill, H. M., Rigosi, A. F. & Heinz, T. F. Population inversion and giant bandgap renormalization in atomically thin $WS_2$ layers. *Nat. Photon.* **9**, 466-470 (2015).

26. Sun, Y. *et al.* 946 nm Nd: YAG double Q-switched laser based on monolayer $WSe_2$ saturable absorber. *Opt. Express* **25**, 21037-21048 (2017).

27. Moody, G. *et al.* Intrinsic homogeneous linewidth and broadening mechanisms of excitons in monolayer transition metal dichalcogenides. *Nat. Commun.* **6**, 8315 (2015).

28. Xu, X., Yao, W., Xiao, D. & Heinz, T. F. Spin and pseudospins in layered transition metal dichalcogenides. *Nat. Phys.* **10**, 343-350 (2014).

29. Limonov, M. F., Rybin, M. V., Poddubny, A. N. & Kivshar, Y. S. Fano resonances in photonics. *Nat. Photon.* **11**, 543-554 (2017).

30. Castellanos-Gomez, A. *et al.* Deterministic transfer of two-dimensional materials by all-dry viscoelastic stamping. *2D Mater.* **1**, 011002 (2014).

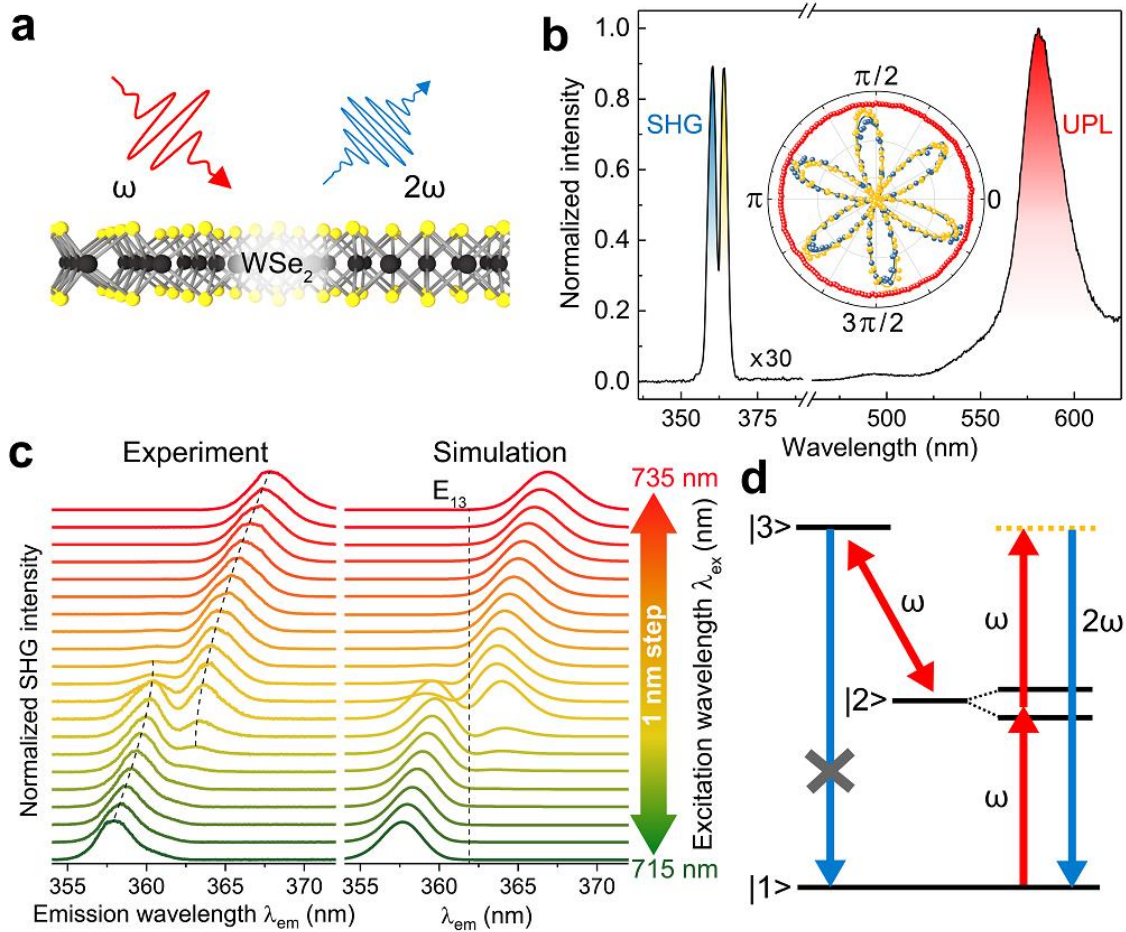

**Figure 1 | Quantum interference in the second-harmonic generation (SHG) of single-layer WSe₂ at 5 K. a**, Illustration of SHG from a single layer of WSe₂, excited by a femtosecond laser. **b**, Typical emission spectrum of monolayer WSe₂ encapsulated by hexagonal boron nitride (hBN) measured under excitation at 724 nm (80 fs pulse length, 80 MHz repetition rate), showing bifurcated SHG centred at 362 nm and an additional upconversion photoluminescence (UPL) feature at 581 nm. The inset plots the intensity of the two SHG peaks (blue and yellow) and the UPL (red) copolarized to the incident laser as the polarization of the driving field is rotated with respect to the crystal axis. **c**, Dependence of the normalized SHG spectrum on pump wavelength in experiment (left) and simulation (right), revealing anticrossing behaviour of the SHG peaks. **d**, Ladder-

type three-level model of resonant SHG and the quantum-interference pathways which lead to spectral suppression. Under coherent drive, the laser field dresses state $|2\rangle$, with transitions $|1\rangle \rightarrow |2\rangle$ and $|3\rangle \rightarrow |2\rangle$ interfering.

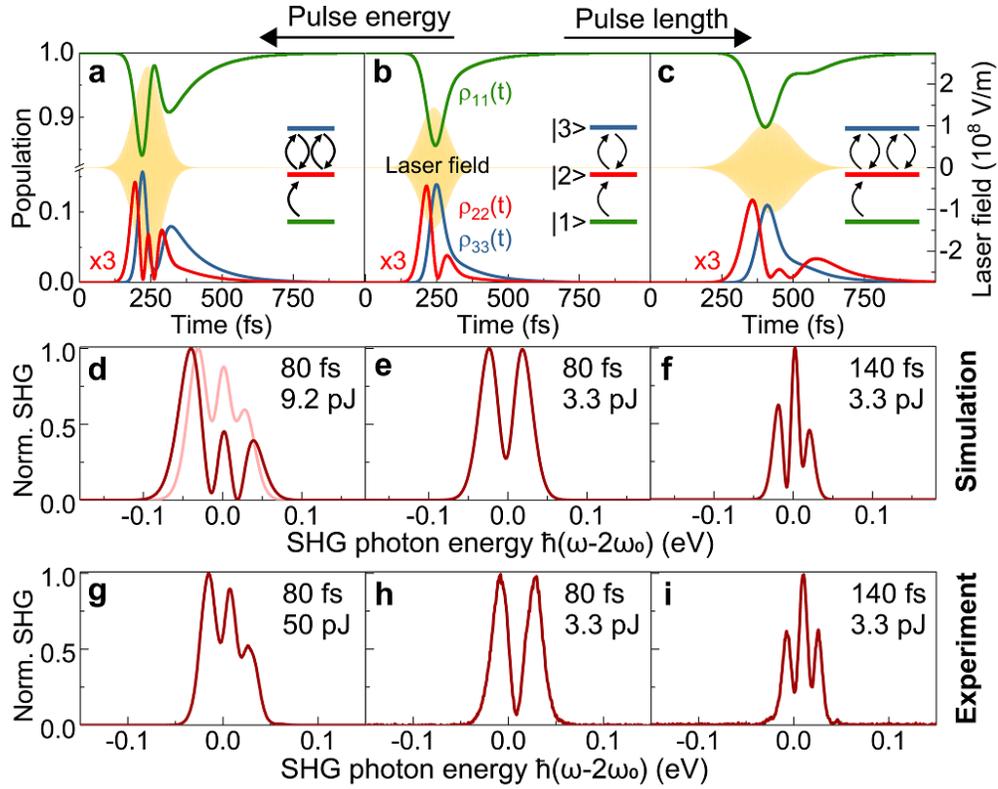

**Figure 2 | Correspondence between Rabi flopping of the strongly driven system and SHG splitting.** Simulated SHG spectra (**d-f**) for the same parameter set (ω₀ pump frequency) and corresponding density matrix dynamics (**a-c**) for different pulse lengths and energies. Rabi flopping between state |2⟩ and |3⟩ under quantum interference is seen in the population dynamics of $\rho_{22}(t)$ and $\rho_{33}(t)$. The number of Rabi flops during the laser pulse (yellow) determines the number of dips in the SHG spectra. **g-i**, Experimental SHG spectra measured on monolayer WSe₂ at 5 K with pulse length and energy of (**g**) 80 fs and 50 pJ, (**h**) 80 fs and 3.3 pJ, and (**i**) 140 fs and 3.3 pJ. The pale-red curve in (**d**) shows a simulation with adjusted parameter set, accounting for many-body interactions, to best fit the experimental spectrum (**g**).

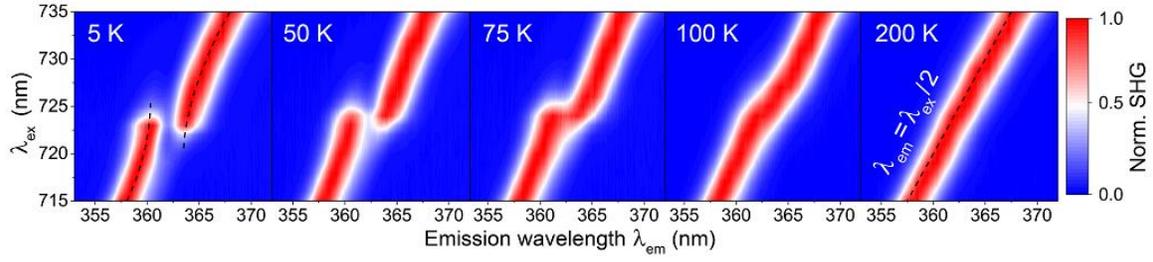

**Figure 3 | Experimental temperature dependence of quantum interference in SHG from hBN-encapsulated monolayer WSe₂.**

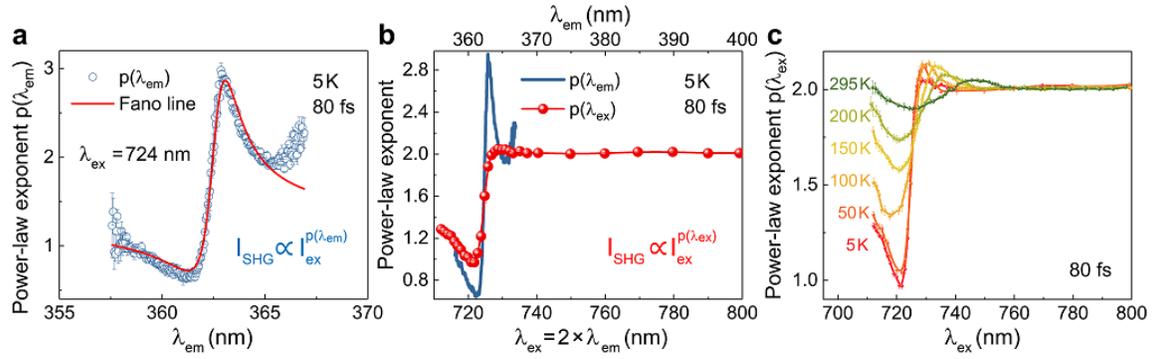

**Figure 4 | Dependence of SHG power-law exponent on excitation and emission wavelength. a**, Spectrally resolved power-law exponent $p(\lambda_{em})$ of the SHG emission from hBN-encapsulated monolayer WSe$_2$, showing a Fano asymmetric line shape. **b**, Comparison of $p(\lambda_{ex})$, the power-law exponent of spectrally integrated SHG intensity as a function of excitation wavelength, and $p(\lambda_{em})$. **c**, Temperature dependence of $p(\lambda_{ex})$. The Fano-like dispersive functionality of $p(\lambda_{ex})$ is retained at room temperature.